\documentclass{elsart}

\usepackage{amssymb}
\usepackage{graphicx}

\begin{document}
\begin{frontmatter}



\title{Dimensional Crossover of Vortex Dynamics Induced by Gd Substitution on Bi2212 Single Crystals\thanksref{ack}}
\thanks[ack]{This work is supported by the National Science Foundation of China
(NSFC 19825111, 10074078) and the Ministry of Science and
Technology of China ( project: NKBRSF-G1999064602 ).}

\author{Z. W. Zhao},
\author{S. L. Li},
\author{H. H. Wen\corauthref{cor}}
\corauth[cor]{Corresponding author.}
\ead{hhwen@aphy.iphy.ac.cn}

\address{National Laboratory for Superconductivity, Institute
of Physics and Center for Condensed Matter Physics, Chinese
Academy of Sciences, P.O. Box 603, Beijing 100080, China }

\author{X. G. Li}
 \address{Department of Material Sciences, University of Science and Technology of China, Anhui 230026, China}
\begin{abstract}
The vortex dynamics of
Bi$_2$Sr$_2$Ca$_{1-x}$Gd$_x$Cu$_2$O$_{8+\delta}$ single crystals
is investigated by magnetic relaxation and hysteresis
measurements. By substituting $Ca$ with $Gd$, it is found that the
interlayer Josephson coupling is weakened and the anisotropy is
increased, which leads to the change of vortex dynamics from 3D
elastic to 2D plastic vortex creep. Moreover, the second
magnetization peak, which can be observed in samples near the
optimal doping, is absent in the strongly underdoped ( with 2D
vortex ) region.
\end{abstract}

\end{frontmatter}

\section{Introduction}
\label{Introduction}
For high temperature superconductors ( HTSC ), the hole
concentration is an essential parameter that controls many
properties. Enormous efforts have been contributed to investigate
the doping dependence of the electronic properties and the
mechanism of high temperature superconductors. While a very
interesting question is that how the vortex dynamics is influenced
by the doping effect.

For highly anisotropic HTSC, such as
Tl and Bi compounds, it has been found that the vortex dynamics
can be driven from 3D to 2D at relatively high fields and high
temperatures\cite{Fisher1,Wen3D2D}. Though it is generally
believed that the doping will strongly affect the coupling between
the Cu-O layers and further influence the vortex dynamics, there
is few such research\cite{Sefrioui}.

In order to see the effect of hole doping on the vortex dynamics,
we have done magnetic relaxation and hysteresis measurements in a
series of Bi$_2$Sr$_2$Ca$_{1-x}$Gd$_x$Cu$_2$O$_{8+\delta}$ single
crystals. By substituting $Ca^{2+}$ with $Gd^{3+}$, the hole
concentration of the samples has been changed from near the
optimal doping to the underdoped regime. It has been found that in
the same series of samples, the substitution strongly reduces the
Josephson coupling between CuO$_2$ layers\cite{Li}. The weaker
interlayer coupling will lead to higher anisotropy. We will
present evidence of the change of vortex dynamics from 3D elastic
to 2D plastic creep by Gd substitution.

\section{Experiment}
The Bi$_2$Sr$_2$Ca$_{1-x}$Gd$_x$Cu$_2$O$_{8+\delta}$ single
crystals with x=0.09, 0.19, 0.32, 0.41 (  named by sample 1, 2, 3,
4 respectively ) were grown by the self-flux method. Details about
the sample preparation were published elsewhere\cite{ZhaoX}. The
in-plane and c-axis resistivity were measured using the standard
four-probe method. Rather sharp superconducting transitions are
observed indicating a good quality of the samples. The magnetic
hysteresis and relaxation measurements were carried out on a
vibrating sample magnetometer ( VSM 8T, Oxford 3001 ) and a
superconducting quantum interference device ( SQUID, Quantum
design, MPMS5.5 ) respectively with the external magnetic field
parallel to c-axis.

To obtain the magnetic relaxation, we first cooled a sample from
above $T_c$ to a desired temperature at zero field ( ZFC ). After
that, the  magnetic field was applied to a high value and then
back to the measured field. The data acquisition ( magnetization
vs time ) started when the field sweep was stopped. It is known
that in the field descending process the vortex escapes from the
sample without facing the geometrical barrier\cite{Yeshurun}. Thus
only the magnetization signal induced by the bulk current was
measured. For the relaxation at each point of T and H, the data
collection lasted for about 3000 seconds.

\section{Results}
\subsection{Doping state and resistivity}
It's well known that a parabolic relationship holds between the
superconducting transition temperature and the hole concentration
$p$\cite{Presland,Tallon}. So the $p$ in the underdoped region can
be determined with $p = 0.16 - [(1-T_c/T_{max})/82.6]^{1/2}$ from
the measured values of $T_c$. Because sample 1 is near the optimal
doping, here we use its transition temperature 87K as $T_{max}$.
The dependence of $T_c$ vs. $p$ of the four crystals is shown in
Fig.1(a). Other three samples are in the underdoped regime with
$T_c$ = 84, 79, 63 K. Samples 3 and 4 are heavily underdoped. From
the inset of Fig.1(a), it can be seen that $T_c$ also shows a
nearly parabolic relationship with the Gd concentration $x$. This
indicates that the Gd substitution influences the hole
concentration and reduces the carrier density. Though it is hard
for us to measure the oxygen content of the four crystals
precisely, by the fact that the c-axis resistivity increases with
the Gd substitution, it is clear that the oxygen content decreases
with $x$\cite{Li2,Watanabe,Villard}.
\begin{figure}
\includegraphics[width=10cm]{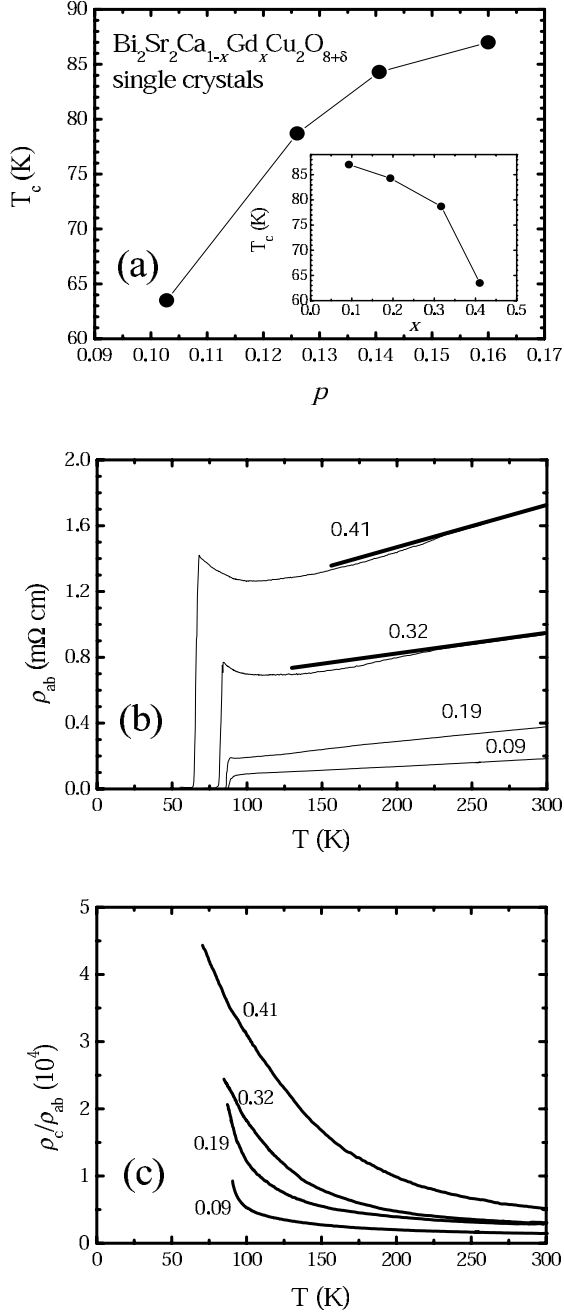}

\caption{(a) Superconducting transition temperature $T_c$ vs. hole
concentration in Bi$_2$Sr$_2$Ca$_{1-x}$Gd$_x$Cu$_2$O$_{8+\delta}$
single crystals. Inset shows the relationship of $T_c$ vs. the
content of Gd. (b) In-plane resistivities of the four samples.
Substitution increases not only the magnitude of resistivity, but
also changes the temperature dependence of the resistivity.
Samples 1 ($x=0.09$) and 2 ($x=0.19$) show a metallic behavior,
but samples 3 ($x=0.32$) and 4 ($x=0.41$) behave more like
semiconductors or insulators. (c) The ratio of c-axis to in-plane
resistivities, from which, the anisotropy can be determined.}
\label{fig1}
\end{figure}

The in-plane resistivity is presented in Fig.1(b). With increasing
the amount of Gd, the normal state resistivity of four samples
varies greatly. Near $T_c$, the resistivity of the lightly
substituted samples 1 and 2 decreases linearly with lowering
temperature and shows metallic behavior. While, for the heavily
substituted samples 3 and 4, the resistivity rises up quickly when
the temperature decreases showing a behavior more like a
semiconductor or insulator. The different temperature dependence of
resistivity between the lightly and heavily substituted samples reflects
the underlying different electronic ground states and may strongly
influence the dissipation process in these samples. In Fig.1(c),
we show the ratio of c-axis to in-plane resistivity. The
anisotropy $\gamma =(m_c/m_{ab})^{1/2} \approx
(\rho_c/\rho_{ab})^{1/2}$ can be estimated from the resistivity
ratio near $T_c$. The enhancement of resistivity ratios indicates
that the anisotropy is enlarged by Gd substitution.

\begin{figure}
\includegraphics[width=13cm]{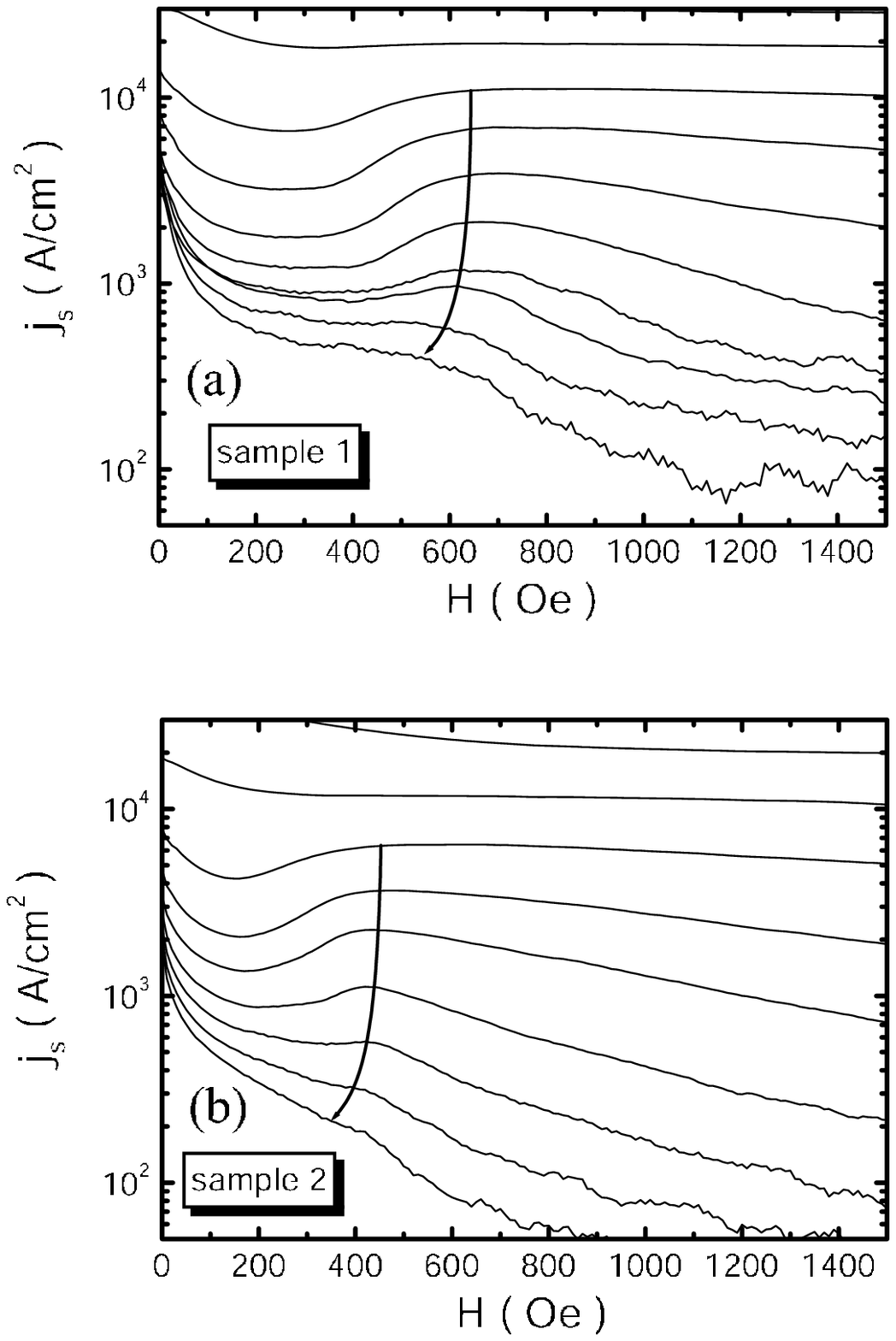}
\caption{ The magnetic field dependence of the $j_s$ in (a) sample
1, from 21 to 45 K and (b) sample 2, from 17 to 42 K with a step
of 3 K. The crossing points of the arrowhead line with the $j_s -
H$ curves indicate the positions of second peaks. }
\end{figure}

\subsection{Second peak}
In samples 1 and 2, a second peak ( SP ) can be observed on the
magnetic hysteresis loop ( MHL ) when temperature is at about 20 -
40 K. However, no second peak is observed in samples 3 and 4 at
any temperatures. Following previous studies, we referred the
second peak in the descending branch of a hysteresis loop as
$H_{sp}$. Because the total magnetic moment M is dependent on the
sample volume and shape, it is better to use the current density
as a basic quantity in comparing the property between different
samples. The current density $j_s$ can be determined by using $j_s
= 20M/Va(1-a/3b)$ based on the extended Bean critical state
model\cite{Higuchi,Bean}, where M is the total magnetic moment of
the sample measured in field descending branch and V, a and b are
the volume, width and length ( $a < b$ ) of the sample,
respectively.

In Fig.2(a) and Fig.2(b), we show the magnetic field dependence of
$j_s$ of samples 1 and 2. The crossing points of the arrowhead
line with the $j_s - H$ curves indicate the positions of second
peaks. At temperatures below about 45 K, the second peak is weakly
dependent on the magnetic field, which in sample 2 is near 450 Oe,
being lower than 600 Oe in sample 1. Moreover, the magnitude of
second peak in sample 2 is also smaller than that in sample 1.

All these facts indicate that with increasing the amount of Gd,
the magnitude of second peak decreases, its position moves to
lower fields and finally disappears completely in the heavily
underdoped samples 3 and 4. Similar results on
Bi$_2$Sr$_2$Ca$_{1-x}$Gd$_x$Cu$_2$O$_{8+\delta}$ single crystals
are reported by other authors\cite{Villard}.

\subsection{Magnetic relaxation}
In Fig.3, the time dependence of the superconducting current
density $j_s$ of sample 1 at 0.1 T is shown at temperatures
ranging from 2 K to 40 K. It is clear that the magnetic relaxation
can be described by a double-logarithmic relation in the low
temperature region.

\begin{figure}
\includegraphics[width=16cm]{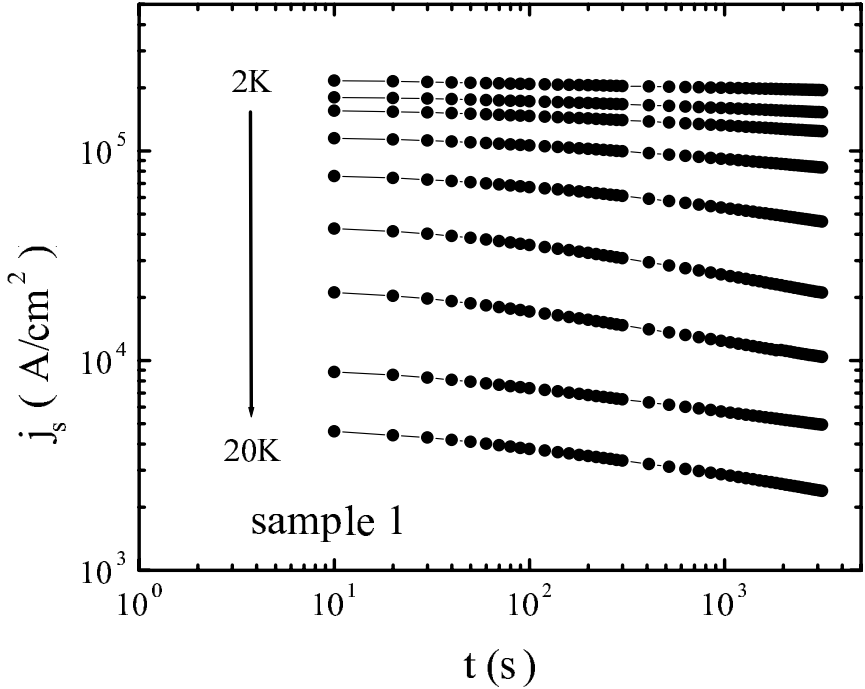}
\caption{ The time dependence of the superconducting current
density $j_s$ of sample 1 at 0.1 T from 2 to 20 K with a step of 2
K. }
\end{figure}

In our experiment, the magnetic relaxation measurements were
carried out at the magnetic fields of 0.5k, 1k, 3k to 5k Oe. The
conventional relaxation rate $S(T)=-j_s^{-1}dj_s/dlnt \approx
-dlnj_s/dlnt$ is showed in Fig.4. To investigate the influence of
doping, the data of four samples is plotted together at the same
magnetic field.

According to the Thermally Activated Flux Motion ( TAFM ) model,
the flux creep should entirely stop at 0 K. However, there is a
residual relaxation rate in the vanishing temperature limit at all
fields for all four samples. This is commonly explained by the
vortex quantum tunnelling creep\cite{Blatter}. For Bi2212 single
crystals, the quantum tunnelling effect is significant below only
about 4 K\cite{Monier}.

When the temperature is increased above 4 K, S(T) appears linear
with T. This is predicted by the Anderson-Kim
Model\cite{Anderson}, which assumes a linear activation energy
$U(j_s) = U_c(T)(1 - j_s/j_c)$. Actually this model is always
valid when $j_s$ is close to the critical current density $j_c$,
since for any kind of $U-j$ relation one can make a Taylor's
expansion with $(1-j/j_c)$ as the variable.

Generally, the linear temperature dependence of S(T) will not
exist at higher temperatures. For samples 1 and 2, when the
temperature increases up to a higher value, a plateau appears at
all of the applied fields. With increasing magnetic fields, the
plateau exhibits an interesting evolution. One can see that at
fields of 0.5k Oe and 1k Oe, the plateau appears at about 14 K and
10 K and extends to some temperature higher than 20 K. But at
higher fields 3k Oe and 5k Oe, the width of the plateau reduces
quickly and the onset points move down to 8 K and 6 K. Clearly, at
a much higher field, the plateau will disappear completely. This
behavior is very similar to what happens in the highly anisotropic
Tl2212 film\cite{Wen3D2D}. However, it differs significantly from
that in the weakly anisotropic YBCO, where the plateau extends
more widely and exists at least up to 7T\cite{Malozemoff1}.

For samples 3 and 4, however, no plateau is observed at any field
used here. Because of the very low irreversibility lines in these
two samples, we are not able to obtain the relaxation rate at
higher temperatures due to small signals and rapid relaxation of
magnetization. The contrast of the existence and nonexistence of
the plateau of relaxation rate S for samples 1, 2 and samples 3, 4
clearly demonstrate a change a vortex dynamics due to the Gd
substitution. In the following, we will show that this difference
is actually correlated with the dimensional transition of vortex
dynamics.

\section{Discussion}
\subsection{Second peak and anisotropy}
The second peak has been interpreted with many models, e.g. effect
of surface barrier\cite{Kopylov}, dimensional
crossover\cite{Vinokur}, vortex and defect matching\cite{Yang}.
For the strongly layered superconductor Bi-2212, it has been
argued that the SP is originated from a phase transition from a
quasi-ordered Bragg glass ( BG ) in low field region to a
disordered vortex glass ( VG ) in high field
region\cite{Giamarchi}. This picture has been supported by many
experiments\cite{Cubitt,Khaykovich,Beek,Gaifullin}. According to
this theory, dense disorders will strongly distort the vortex
lattice leading to the disappearance of the Bragg glass state.
Meanwhile the transition / crossover between a Bragg glass and a
vortex glass is predicted for a 3D or quasi-2D vortex system. When
the anisotropy is increased, the disappearance of the SP is
expected\cite{Zeng}.

In our present experiment, the Gd substitution introduces more
defects into the samples and also increases the anisotropy. Both
will lead to the disappearance of SP. This has been indeed
observed in the experiment. However, from present study, it is
hard to judge whether the disappearance of SP in the heavily
underdoped region is due to the increase of disorders or the
vortex system changing from 3D to 2D, or by both.

\subsection{The plateau of S-T}
The plateau on the $S  vs. T$ curves has been found in a variety
of HTSC. Commonly it is ascribed to a nonlinear $U-j$ relationship
predicted by the Vortex Glass ( VG )\cite{Fisher2} and Collective
Pinning ( CP ) models\cite{Blatter,Feigelman}. According to both
theories, the activation barrier U depends on the current density
$j_s$ as $U(j_s)\propto 1/j_s^\mu$, where $\mu$ is the glassy
exponent. This relation naturally explains the divergence of $U$
in the small current limit of $j_s$ and predicts a state with zero
linear resistivity. Later on, an interpolating expression was
proposed by Malozemoff {\it et al.}\cite{Malozemoff2}

\begin{equation}
U(j_s, T) = \frac{U_c}{\mu} [(\frac{j_c}{j_s})^{\mu}-1]
\label{Ujelastic}
\end{equation}
where $U_c$ is the characteristic pinning energy and $j_c$ is the
critical current density. The $\mu, U_c, j_c, j_s$ are dependent
on temperature and external field. This relation is rather general
in respect that it describes different $U-j$ relations based on
various models.

In the conventional magnetic relaxation process, $dj_s/dt$ is
assumed to be proportional to the electric field E induced by the
TAFM

\begin{equation}
\frac{dj_s}{dt} \propto E = v_0Bexp(-\frac{U}{k_BT})
\label{Ej}
\end{equation}
Here $v_0$ is the attempting velocity of the flux motion.

Combining the eq.(\ref{Ujelastic}), eq.(\ref{Ej}) and the
definition of relaxation rate $S(T) = -dlnj_s/dlnt$, a simple
relation was derived by Wen{\it et al.}\cite{Wen95PhysicaC}
\begin{equation}
S(T) = \frac{T}{U_c/k_B + \mu CT}
\label{S}
\end{equation}
where $C = - (dlnj_s/dT)(T/S)|_{T=0} = lnv_0B/lnE$ is a parameter
independent of T. For four samples studied here, the value of C
estimated by Maley's method is about 10 to 15\cite{Wen3D2D,Maley}.

From eq.(\ref{S}), it can be seen that the shape of $S  vs. T$
curve is mainly determined by two terms in the denominator. When
$U_c/k_B$ is much larger than $\mu CT$, the eq.(\ref{S}) produces
the linear temperature dependence. While, if $\mu CT$ prevails
over $U_c/k_B$, this equation will give a result as

\begin{equation}
S(T) = \frac{1}{\mu C}
\label{Splateau}
\end{equation}

If $\mu$ is not strongly dependent on T, this formula predicts
that S(T) will have a plateau, just like what appears on the $S  vs. T$ curves of samples 1 and 2.

The ability to explain the plateau in S(T) found from experiment has been regarded as a
major success of the VG or CP theory\cite{Yeshurun}. Now, let us
concentrate on the detailed character of the plateau.

(i){\it The sharp crossover from a linear dependence to a plateau
with increasing T}. ----- From Fig.4, it can be seen that the
crossover from a linear dependence to a plateau is rather sudden,
especially at lower fields, 0.5k and 1 k Oe. Since a plateau will
only appear when $\mu CT$ exceeds $U_c/k_B$, the sharp transition
may manifest that there is a rapid variation of either $\mu, C$ or
$U_c$. Considering $C$ is independent of T as stated above, and
 $U_c$ is weakly dependent on temperature in the
temperature region studied here\cite{Wen3D2D}, the only
possibility is that $\mu$ rises suddenly at a certain temperature.

In the CP model, $\mu$ is predicted to have different values in
various creep regions\cite{Blatter}: it is 1/7 in the single
vortex creep region, and with increasing temperature it will
become 3/2 in the small bundles creep region and then 7/9 in the
large bundles region. From 1/7 to 3/2, the $\mu$ enlarges about 10
times. Such a variation is large enough to result in a sudden rise
of relaxation rate. So it is reasonable to regard the crossover
from a linear dependence to a plateau as a proof of the vortex
dynamics transferring from the single vortex to the small bundles
creep region. Furthermore, when the vortex dynamics transfers from
the small to large bundle pinning region, $\mu$ will vary between
3/2 and 7/9, which gives rise to a relatively stable value of S,
being consistent with the existence of a plateau. Taking $\mu =
7/9$ and $C=13$, eq.(\ref{Splateau}) gives S$_{plateau}$ = 10\%,
which is near the value found in our experiment. All these
analyses above are based on 3D elastic flux
dynamics\cite{Blatter}. For 2D in the high field region, the
pinning to pancake vortex is weak and it can easily jump out of
the pinning wells. The relaxation rate rises quickly with
increasing temperature and field. Therefore, there will be no
plateau in the $S-T$ curves just like what happens to samples 3
and 4.

\begin{figure}[t]
\includegraphics[width=16cm]{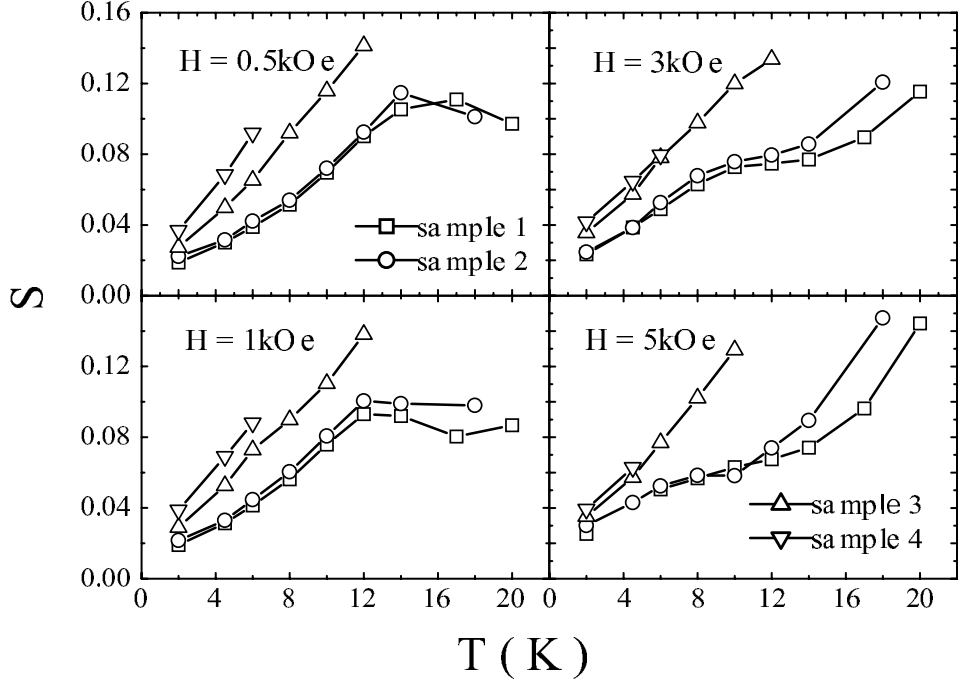}
\caption{ The relaxation rate S(T) {\it vs.} T at the fields of
0.5k, 1k, 3k to 5k Oe. At each magnetic field, the results of four
samples with different Gd concentration are showed together. For
samples 1 and 2, a clear plateau appears. Its onset point moves
down to lower temperatures and the width shrinks with increasing
field. For samples 3 and 4, the S rises much faster and there is
no plateau. }
\end{figure}

(ii) {\it The variation of  onset point and width of a plateau
with increasing fields.}-----  One can easily see that the onset
point of a plateau moves down to a lower temperature with
increasing field. A straightforward explanation is that
increasing field will lead to the reduction of $U_c$, so the
crossover to a plateau occurs at a lower temperature (see from
eq(3)). It is known that increasing the magnetic field tends to
suppress the coupling between superconducting layers. The Gd
substitution can also weaken the interlayer coupling. Thus it can
be expected that $U_C$ will decrease with more substitution.

In addition, the width of the plateau decreases at higher fields.
That indicates that the ending point of a plateau moves down to a
lower temperature with increasing field. As soon as the plateau
ends at a certain temperature, the S(T) begins to rise quickly
toward the irreversibility line. This drastic rise can be
understood as due to the plastic vortex creep, which will be
discussed in next subsection.

\subsection{$U_c$ and plastic creep}
The moving vortices in type-II superconductors are under the
action of Lorentz force and pinning force. According to the CP
theory, to balance the elastic and pinning energy, the
collectively pinned vortex ( bundle ) has a length $L_c$ with the
collective pinning energy $U_c$\cite{Blatter}
\begin{equation}
L_c \simeq \gamma^{-1}\xi (j_0/j_c)^{1/2}
\label{Lc}
\end{equation}
\begin{equation}
U_c \simeq \gamma^{-1}\xi ln\kappa (\Phi_0/4\pi \lambda)^2 (j_c/j_0)^{1/2}
\label{Uc}
\end{equation}
where $j_0 \approx \Phi_0 c/12 \pi \lambda^2 \xi$ is the depairing
current density, L$_c$ and U$_c$ are both inversely proportional
to the anisotropy $\gamma$.

\begin{figure}
\includegraphics[width=16cm]{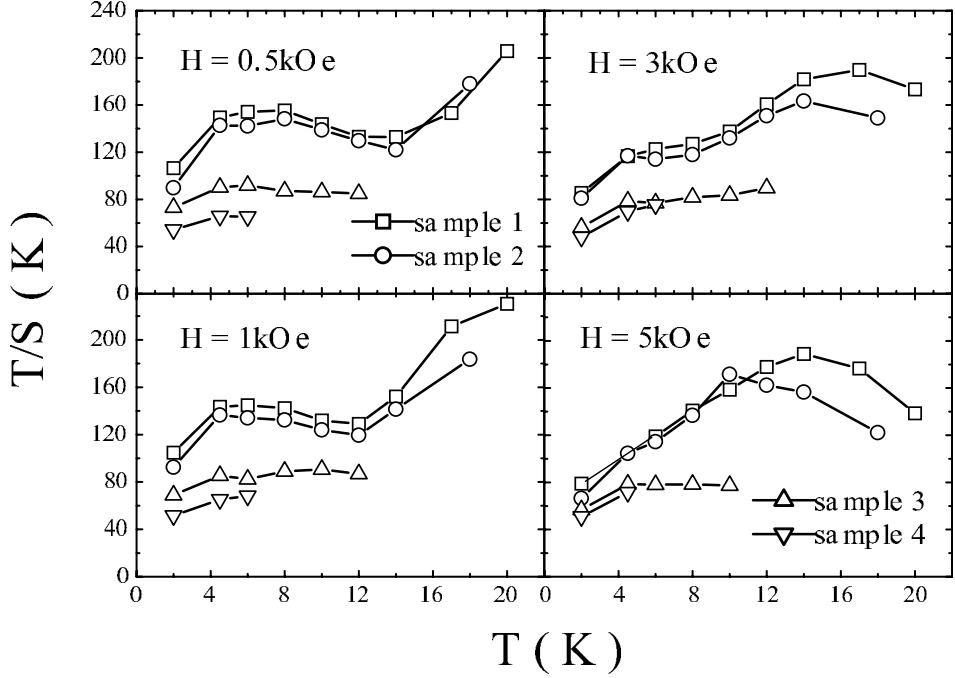}
\caption{ The temperature dependence of T/S with similar
arrangement of Fig.3. $U_c = T/S|_{T=0}$ can be determined by
extrapolating the curves to 0 K. }
\end{figure}

To determine $U_c$ from relaxation data, we rewrite eq.(\ref{S})
in the form
\begin{equation}
\frac{T}{S} = U_c/k_B + \mu CT
\label{TS}
\end{equation}

From this equation, one can see that $T/S|_{T=0}= U_c(T=0)/k_B $.
Since $U_c$ is weakly temperature dependent when $T<<T_{irr}$ in
highly anisotropic superconductors\cite{Wen3D2D}, we can get the
general $U_c = U_c(0)$ from the low temperature limit of T/S.

The T/S {\it vs.} T curves are showed in Fig.5 with the same
arrangement in Fig.4. Due to vortex quantum tunnelling creep, S
has a residual value in the vanishing temperature. Thus T/S begins
to drop to zero when the quantum effect dominates. This crossover
occurs in our samples is at about 4 K. To determine the $U_c(0)$,
we extrapolate the curve from above 4 K to 0 K. Being limited by
few data points, we can only obtain some semi-quantitative values
of $U_c$. These data, however, can still show the underlying
characters of vortex dynamics induced by the Gd substitution.

\begin{figure}
\includegraphics[width=16cm]{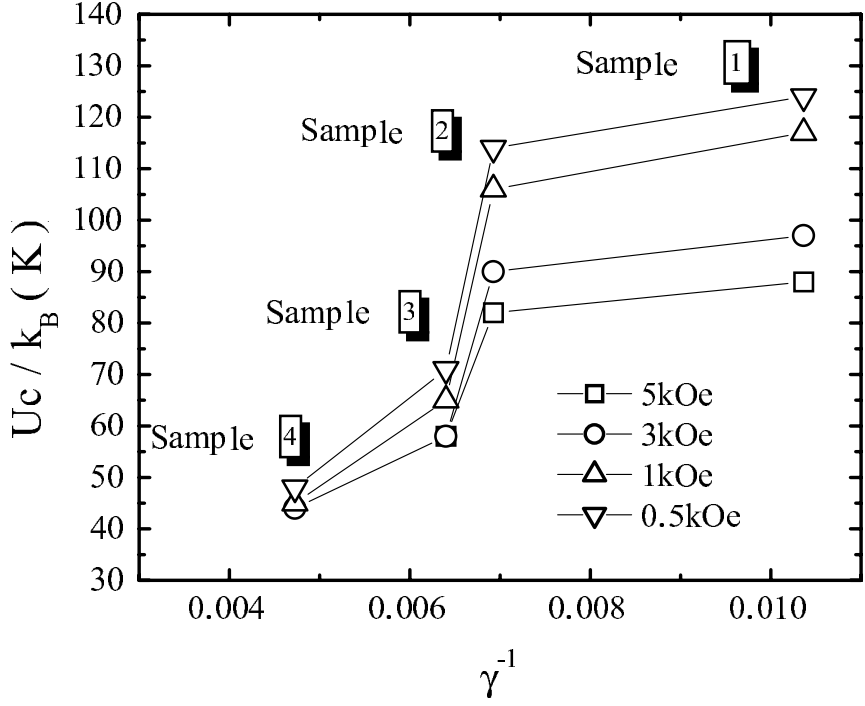}
\caption{ The collective pinning energy U$_c$ {\it vs.}
$\gamma^{-1}$ at different fields. For samples 3 and 4, the
influence of the magnetic field on U$_c$ is smaller than that for
sample 1 and 2. With more Gd substitution, the U$_c$ decreases to
about a half from sample 1 to sample 4. }
\end{figure}

The relation of $U_c(0)$ vs. anisotropy $\gamma$ is presented in
Fig.6. For samples 1 and 2, it is clear that $U_c$ decreases with
increasing field. This accords with the fact that the onset point
of a plateau moves down to lower temperatures at higher fields.
While, for samples 3 and 4, $U_c$ only weakly depends on magnetic
fields, that coincides with 2D character of flux dynamics in these
heavily underdoped samples. Another remarkable feature in Fig.6 is
that $U_c$ increases with $\gamma^{-1}$, just as what the CP
theory predicts. Also according to the CP, the decrease of $U_c$
with increasing fields and dopants can only be induced by reducing
the collective pinning length $L_c$. So this is another proof that
the vortex dynamics in heavily underdoped sample 3 and 4 tends to
be 2D.

Besides $U_c$, from eq.(\ref{TS}) one can also qualitatively
determine $\mu$, at least for its sign. Since the $C$ is positive,
the $\mu$ will be positive when T/S is larger than $U_c/k_B$ and
be negative when T/S is less than $U_c/k_B$. For the VG and CP
theories, the elastic flux motion is expected and $\mu$ must be
positive. For a negative $\mu$, eq.(\ref{Ujelastic}) can be
written in the form

\begin{equation}
U(j_s) = U_0 [1- (\frac{j_s}{j_c})^{\alpha}]
\label{Ujplastic}
\end{equation}
where $U_0=-U_c/\mu$ and $\alpha=-\mu$ are both positive.
Eq.(\ref{Ujplastic}) predicts that $U$ will not diverge when $j_s$
approaches zero. This indicates that there is always a finite
linear resistivity which is a character of the dislocation
mediated plastic motion\cite{Wen3D2D,vanBeek}. It is different
from the situation in the elastic flux motion, where $\mu$ is
positive and the $U$ will diverge when $j_s$ approaches zero.

From Fig.5, no negative $\mu$ was found at all temperatures and
fields investigated here. However, the trend of $\mu$ transferring
from positive to negative is obvious. For sample 1 and 2, at the
fields of 3k and 5k Oe, the T/S begins to decrease quickly after a
rise. The crossover corresponds to the ending point of the plateau
on S vs T curves. So the fast rising of S after a plateau can be
understood as the vortex system entering the region dominated by
plastic vortex creep. For samples 3 and 4, T/S is stable and
doesn't drop in the measured narrow temperature range. But, in
fact because the irreversibility temperatures of sample 3 and 4
are very low, S will take a fast rise at higher temperatures. Thus
it can be expected that T/S of sample 3 and 4 will drop quickly
with further increasing temperatures. Therefore, we can conclude
that the plastic vortex motion not only tends to occur at high
temperatures and high fields for samples 1 and 2 but also should
be more pronounced in highly anisotropic systems like samples 3
and 4. In our experiment, it is the Gd substitution that
increases the anisotropy and leads to the plastic flux motion.

\section{Concluding remarks}
In summary, we have investigated the vortex dynamics in a series
of  Bi$_2$Sr$_2$Ca$_{1-x}$Gd$_x$Cu$_2$O$_{8+\delta}$ single
crystals. By weakening the interlayer coupling and increasing the
anisotropy, the Gd substitution has significantly influenced the
vortex dynamics of Bi2212. The second peak is only observed near
the optimal doping ( samples 1 and 2 ), but in the heavily
underdoped region ( samples 3 and 4 ), the increase of disorders
and/or anisotropy causes the second peak to disappear. Moreover,
with increasing Gd concentration, the plateau of S-T disappears
and the collective pinning energy $U_c$ decreases, which indicates
that the vortex system changes from 3D to 2D and the vortex tends
to creep plastically.



\end{document}